\documentclass[twocolumn,aps,prl,showpacs,preprintnumbers,amssymb,longbibliography]{revtex4-1}

\usepackage{epsfig}
\usepackage{dcolumn}
\usepackage{bm}
\usepackage{hyperref}

\usepackage{amsmath}
\usepackage{amssymb}
\usepackage{latexsym}
\usepackage{epsfig}
\usepackage{color}
\usepackage{mathtools}
\usepackage{youngtab}
\usepackage{bigints}
\usepackage{sidecap}

\def\z2z2{$\IC^3/(\IZ_2\times\IZ_2)$}

\def\id{{\bf 1}}

\def\a{\alpha}

\def\b{\beta}

\def\d{\delta}\def\D{\Delta}

\def\h{\eta}
\def\k{\kappa}
\def\l{\lambda}

\def\p{\pi}

\def\s{\sigma}

\def\th{\theta}

\def\beq{\begin{equation}}\def\eeq{\end{equation}}
\def\beqa{\begin{eqnarray}}\def\eeqa{\end{eqnarray}}
\def\barr{\begin{array}}\def\earr{\end{array}}

\def\wt{\widetilde}

\def\ds {{\del \hspace{-6.4pt} \slash}\;}

 \let\br=\bigr

\def\bd{\begin{document}}
\def\ed{\end{document}}
\def\ba{\begin{array}}
\def\ea{\end{array}}
\def\bea{\begin{eqnarray}}
\def\eea{\end{eqnarray}}
\def\ft#1#2{{\textstyle{{\scriptstyle #1}\over {\scriptstyle #2}}}}
\def\fft#1#2{{#1 \over #2}}
\newcommand{\be}{\begin{equation}}
\newcommand{\ee}{\end{equation}}
\newcommand{\eq}[1]{(\ref{#1})}
\def\eqs#1#2{(``{#1}-``{#2})}
\def\det{{\rm det\,}}
\def\tr{{\rm tr}}
\newcommand{\ho}[1]{$\, ^{#1}$}
\newcommand{\hoch}[1]{$\, ^{#1}$}
\def\ra{\rightarrow}

\def\Xh{\hat{X}}
\def\ah{\hat{a}}
\def\xh{\hat{x}}
\def\yh{\hat{y}}
\def\ph{\hat{p}}
\def\G{{\cal G}}
\def\Dth{{\Delta_\th}}

\def\bk{{\bf k}}
\def\bx{{\bf x}}
\def\br{{\bf r}}
\def\tr{{\rm tr \,}}
\def\Tr{{\rm Tr \,}}
\def\diag{{\rm diag \,}}
\def\tg{{\rm tg \,}}
\def\ov{\overline}

\def\preal{{\rm Re\,}}
\def\pim{{\rm Im\,}}

\def\ds{\displaystyle}
\def\yzero{\smash{\hbox{$y\kern-4pt\raise1pt\hbox{${}^\circ$}$}}}
\def\p{\partial}
\def\a{\alpha}
\def\b{\beta}
\def\g{\gamma}
\def\d{\delta}
\def\beq{\begin{equation}}
\def\eeq{\end{equation}}
\def\beqa{\begin{eqnarray}}
\def\eeqa{\end{eqnarray}}
\def\Om{\Omega}
\def\om{\omega}
\def\th{\theta}
\def\vt{\vartheta}
\def\vphi{\varphi}
\def\-{\hphantom{-}}
\def\ov{\overline}
\def\s2{\frac{1}{\sqrt2}}
\def\wh{\widehat}
\def\wt{\widetilde}
\def\oh{\frac{1}{2}}
\def\tr{{\rm tr \,}}
\def\Tr{{\rm Tr \,}}
\def\diag{{\rm diag \,}}
\def\vac{|0 \rangle}
\def\vm{\relax{n_{\text{v}}}}
\def\tv{\tilde v}
\def\Dsl{\,\raise.15ex\hbox{/}\mkern-13.5mu D} 
\def\id{{\rm 1}}

\def\ti{\times}
\def\til{\tilde}
\def\eps{\epsilon}
\def\k{\kappa}
\def\A{\Arrowvert}
\def\cw{{\cal W}}
\def\G{\Gamma}
\def\car{{\cal R}}
\def\l{\lambda}
\def\raw{\rightarrow}
\def\Raw{\Rightarrow}
\def\inte{{\bf Z}}
\def\cpx{{\bf C}}
\def\real{{\bf R}}
\def\Lam{\Lambda}
\def\D{\Delta}
\def\cb{{\cal B}}
\def\ca{{\cal A}}

\def\re{\mbox{Re }}
\def\im{\mbox{Im }}
\def\tr{\mbox{Tr}}
\def\str{\mbox{STr}}

\def\IC{\mathbb{C}}
\def\IN{\mathbb{N}}
\def\IZ{\mathbb{Z}}
\def\IR{\mathbb{R}}
\def\IP{\mathbb{P}}
\def\Id{{\mathbb{I}}}

\begin{document}

\preprint{MAD-TH-18-03}

\title{Mass Hierarchies and Dynamical Field Range}
\author{Aitor Landete and Gary Shiu}
\affiliation{\small\slshape   Department of Physics, University of Wisconsin-Madison, Madison, WI 53706, U.S.A }
\begin{abstract}
Several swampland conjectures suggest that there is a critical field range beyond which the effective field theory (EFT) description breaks down in quantum gravity. In applications of these conjectures, however, the field range of interest is the field space distance traced by the physical trajectory that solves the equations of motion. We refer to this field space distance as the dynamical field range.  We show that in the absence of a mass hierarchy between the light and heavy fields, the trajectory of the light field does not, in general, follow a geodesic in field space. Then, stabilizing the heavy fields at the minimum of their potential does not accurately describe the dynamics of the light field in general. A mass hierarchy can delay the breakdown of the EFT, and extend the effective field range. We illustrate these subtleties of multi-field dynamics with axions in Type II string compactifications.

\end{abstract}
\maketitle

\section{Introduction}

In recent years, an increasing amount of evidence suggests that not every low-energy EFT can be consistently coupled to quantum gravity.
The quantum gravity constraints that have been unveiled so far have many ramifications. 
For instance, studies of the generic properties of 
quantum gravity point to a restriction on the allowed field range in ultraviolet (UV) complete theories. 
This restriction takes various
forms. The weak gravity conjecture \cite{ArkaniHamed:2006dz} asserts that for every long range force, there exists a state whose charge-to-mass ratio is bigger than that of an extremal black hole. This implies upon dualizing an axion-like field to a $U(1)$ gauge field 
that the axionic decay constant, $f$, is limited by $f \cdot S_{\rm inst} < {\cal O} (1) M_P$  \cite{Brown:2015iha, Brown:2015lia} where $S_{\rm inst}$ is the instanton action that controls the non-perturbative breaking of the shift symmetry. A more general statement about field ranges, known as the Swampland Distance Conjecture (SDC), was put forth in \cite{Ooguri:2006in}.
This conjecture is based on the observation that in known string constructions, a tower of states become exponentially light as we traverse a large distance $d(p_0, p)$ in field space:
\begin{equation}
M \sim M_0 e^{-\lambda d(p,p_o)}~,
 \label{swampconj}
\end{equation}
where $\lambda$ is some unspecified positive constant, and thus the EFT breaks down beyond a critical distance $d (p_0,p) > \lambda^{-1}$. For recent considerations about this conjecture we refer the reader to \cite{Grimm:2018ohb,Heidenreich:2018kpg}.

The conjectured restrictions on the allowed field range, if proven, have wide-ranging implications, one of which concerns the amplitude of 
gravitational waves generated by inflation. A kinematic bound due to Lyth \cite{Lyth:1996im} relates observable tensor modes to super-Planckian 
field displacements:
\begin{equation}
\frac{\Delta \phi}{M_{Pl}}  \gtrsim \mathcal{O}\left(1\right) \times \sqrt{\frac{r}{0.01}}~,
\end{equation}
where $r$ is the tensor-to-scalar ratio. Thus, if quantum gravity can impose a strict upper bound on the inflaton field range, one may gain a better certainty on our target for $r$. Besides the challenge in maintaining control of the EFT over a super-Planckian field displacement, a detectable level of tensor modes also poses an additional challenge on the UV completion of inflation. This is because for large-field inflation (say $r \gtrsim 10^{-2}$), the Hubble scale during inflation is $H \sim 10^{14}$ GeV\cite{Ade:2015lrj}. As UV complete theories of gravity typically involve new degrees of freedom below the scale of quantum gravity, 
it is a formidable task to pack the associated energy scales in between $H$ and $M_{\text{Pl}}$. To be concrete, constructing single-field inflation models from string theory requires a hierarchy of scales:
\be
M_{\varphi} < H < M_{\text{moduli}} < M_{\text{KK}} < M_{s} < M_{\text{Pl}}\, ,
\label{hierarchies}
\ee
where $M_{\varphi}$ is the inflaton mass, $M_{\text{moduli}}$ denotes generically the mass scale of the stabilized moduli \footnote{There can be further hierarchies within the moduli sector, e.g., in Type IIB flux compacitications, the Kahler moduli can be lighter than the complex structure moduli. Here, we assume that the moduli are stabilized within the 4D EFT, and thus $M_{\text{moduli}} < M_{KK}$.}, $M_{KK}$ refers to the mass of the lightest Kaluza-Klein replica, and $M_s$ is the string scale. The challenge in maintaining this hierarchy is thus two-fold: 1) the moduli stabilization mechanism should generate this separation of scales (see, e.g., \cite{Blumenhagen:2014nba} for a discussion of this issue in flux compactifications),  and 2) backreaction of the heavy fields (which in general has the effect of flattening the inflaton potential \cite{Dong:2010in,Buchmuller:2015oma})
throughout the inflationary trajectory  should be consistently taken into account. 

Given the
above considerations,
it is important to define a proper measure of the maximum field range. The main goal of this letter is to make clear the distinction between
kinematic and dynamical field ranges.
While the SDC is a kinematic statement about the moduli space, i.e., it tracks the change in the low-energy EFT from $p_0$ to $p$, it makes no reference to the path connecting them being the true trajectory dictated by the underlying dynamics.
In applying the swampland conjectures (e.g., to inflation), however, we are interested in the restriction on the {\it dynamical} field range traversed by the true trajectory that solves the equations of motion (eom). 
As we will see, the dynamical field range and a geodesic distance traversed in field space by the light field only coincide if the light and heavy sectors evolve separately, or when the heavy fields are infinitely heavy. The two field distances deviate from each other precisely because of the hierarchy in Eq.~(\ref{hierarchies}), which mandates the heavy fields to be not arbitrarily heavier than the light field of interest. The trajectory can be approximated by a geodesic if there is a sufficiently large mass hierarchy. Our findings do not contradict the SDC \cite{Ooguri:2006in}, which concerns with the asymptotical behavior as $d( p_0,p) \rightarrow \infty$.
However, what we found has bearings on the Refined Swampland Distance Conjecture  (RSDC) \cite{Klaewer:2016kiy} which argued that $\lambda \sim {\cal O} (1)$.  We will show explicitly that the constraints on field ranges imposed when $\lambda \sim {\cal O} (1)$  for massive axions are obtained from unphysical trajectories. This prescription will agree with the dynamical field range in the presence of a sufficiently large mass hierarchy between the light and heavy sectors, but then the critical field range is also extended. We illustrate these subtleties of multi-field dynamics with axions in a 4d $\mathcal{N}=1$ supergravity description of type II compactifications.
Our discussions apply to both Type IIA and Type IIB settings, irrespective of whether the light axion comes from the open-string or the closed-string  sector.

 \section{The dynamical field range}

We now analyze carefully the impact of the kinetic terms of the stabilized fields and discuss under what conditions the trajectory is a geodesic in field space.

\subsection{The Refined Swampland Conjecture for axions}

Before we present a more critical assessment of the dynamical field range, let us revisit the RSDC constraints for axions \cite{Baume:2016psm} which claims that strong backreaction of the closed string sector on the kinetic term of an axion constrains its available field range to $\Delta \phi < {\cal O} (1)  M_{\text{Pl}}$. This claim, if true, excludes the possibility of large-field inflation based on axions, including axion monodromy models \cite{Silverstein:2008sg,McAllister:2008hb} (and its realizations via F-term potential in supergravity \cite{Marchesano:2014mla,Blumenhagen:2014gta,Hebecker:2014eua}).

The approach used in \cite{Baume:2016psm}
to impose constraints on the available field range (in particular, in type IIA compactifications) consists of analyzing the trajectory traced by the minima
of the heavy fields as we vary an axion, $\varphi$. 
Then, one should keep all the fields in the action and adjust their values appropriately. This approach is claimed to be a good approximation irrespective of the existence of a mass hierarchy between the sectors. More concretely, for a massive axionic direction defined as a linear combination of certain fields, $\varphi = h^{i}\nu^{i}$, where $h^{i}$ are constants, it is argued that the field range
of interest is
%
\be
\Delta \phi = \int_{\gamma} \left(h_{i}G^{ij}\left(\varphi\right)h_{j}\right)^{-1/2} d\varphi := \int_{\gamma}\sqrt{G_{\varphi \varphi} \left(\varphi\right)}d\varphi\, ,
\label{defrange2}
\ee
where $G^{ij}\left(\varphi\right)$ is the inverse of the field space metric and $G_{\varphi \varphi}\left(\varphi\right)$ is defined as the kinetic term of the axionic field evaluated along the aforementioned trajectory. One immediately observes that this definition, implicitly, neglects
the displacements of the fields not included in the definition of $\varphi$.
This assumption implies, by default, that the trajectory considered is a geodesic in field space. Then, strong backreaction of the closed-string sector on the kinetic term of $\varphi$ implies that the field range \eqref{defrange2} grows at best logarithmically, i.e. $\Delta \phi \approx \alpha^{-1}\log\left(\varphi\right)$. This is, precisely, the scaling suggested by the SDC. The critical value for the canonically normalized field before the onset of the logarithmic behavior is $\hat{\varphi}_{\text{back}}\sim \alpha^{-1}$. Along the text, hatted variables denote canonically normalized fields. Then, one observes that a tower of states becomes exponentially light as $\varphi$ is displaced, in agreement with the SDC \eqref{swampconj}, if we identify $\alpha \approx \lambda$.
It was argued in \cite{Baume:2016psm} that $\lambda \sim \mathcal{O}\left(1\right)$. Afterwards, in \cite{Valenzuela:2016yny,Blumenhagen:2017cxt} this prescription was generalized to include the open-string sector. There, it was argued that $\lambda$ is proportional to the mass hierarchy between axion, $\varphi$, and the closed-string sector.
%
%



\subsection{Proper field range and validity of the approximation}
%
%
%
We can try to improve the naive estimate of the allowed field range in eq.~\eqref{defrange2} by including the kinetic terms of the heavy fields that were previously neglected. We will see under what conditions can the naive estimate be trusted.

The dynamical field range is the one measured along the one-dimensional trajectory defined by the solution of the eoms. In a multi-field scenario, the homogeneous background fields, $\phi_{0}^{a} = \phi_{0}^{a}\left(t\right)$, in a spatially flat FLRW spacetime, satisfy
\be
D_{t}\dot{\phi}_{0}^{a} + 3H\dot{\phi}_{0}^{a} + V^{a} = 0 \, ,
\label{eoms}
\ee
where the covariant derivative in field space is defined by $D_{t}X^{a} = \partial_{t}X^{a} + \Gamma^{a}_{bc}\dot{\phi}_{0}^{b}X^{c}$ and $ \Gamma^{a}_{bc}$ are the Christoffel symbols derived from the field space metric $G_{ab}$. Here, latin indices $a, b$ denote real coordinates in field space. One may consider the background fields $\phi^{a}_{0}\left(t\right)$ as coordinates of the moduli space describing a curve parametrized by $t$. It is well-known that the solutions to the eoms are tied to two fundamental properties of the system: 1) The existence of mass hierarchies between different sectors and 2) The geometry of the field space metric (we refer the reader to \cite{Rubin:2001in,Tolley:2009fg,Achucarro:2010jv,Shiu:2011qw, Achucarro:2012yr,Gao:2012uq} for more details). The variation of the scalar fields along the trajectory is defined as $\dot{\phi}_{0}^{2}  =G_{ab}\dot{\phi}_{0}^{a}\dot{\phi}_{0}^{b}$. Then, the proper distance traversed along the path parametrized by $t$ will be given by
\be
\Delta \phi = \int \dot{\phi}_{0} dt = \sqrt{G_{ab}\dot{\phi}_{0}^{a}\dot{\phi}_{0}^{b}} dt\, .
\label{realfieldrange2}
\ee
%

To quantify the error of the naive estimate in eq.~\eqref{defrange2}, we 
substitute the solutions which minimize the scalar potential in terms of the displaced field, $\overline{\phi}^{\alpha}\left(\varphi \left(t\right)\right)$, into the action to obtain the proper field range \eqref{realfieldrange2} \footnote{We assume, $\overline{\phi}^{\alpha}\left(0\right) = \overline{\phi}^{\alpha}_{0} > 0$, when $ \overline{\phi}_{0}$ is related with the volumes of cycles of the compactification.}. Note that greek indices run over all the fields in the theory except $\varphi$.
We emphasize here that there is no guarantee
the trajectory given by this procedure is the one
that minimizes the action
 since it is not granted that the eoms are solved. 
Nonetheless, this analysis serves as a self-consistent check of whether the kinetic terms for  $\overline{\phi}^{\alpha}$ can be neglected. 
We have also checked numerically that evaluating 
eq.~(\ref{realfieldrange2}) with the actual solution to the equations of motion does not change qualitatively our conclusions.

In the presence of strong backreaction,
the kinetic terms of the stabilized fields will not be negligible in general. Indeed, we observe that $\dot{\overline{\phi}}^{\alpha}\left(\varphi \left(t\right)\right) \approx \frac{\partial \overline{\phi}^{\alpha}\left(\varphi\right)}{\partial \varphi} \dot{\varphi}$ which implies a proper field range \eqref{realfieldrange2}, assuming no kinetic mixing:
%
\be
\Delta \phi = \int\sqrt{G_{\varphi \varphi}\left(\varphi\right)  + G_{\alpha \beta}\left(\varphi\right)\frac{\partial\overline{\phi}^{\alpha}}{\partial \varphi}\frac{\partial\overline{\phi}^{\beta}}{\partial \varphi}}\dot{\varphi}dt \, ,
\label{rangebetter}
\ee
%
Note that the term $G_{\alpha \beta}\left(\varphi\right)\frac{\partial\overline{\phi}^{\alpha}}{\partial \varphi}\frac{\partial\overline{\phi}^{\beta}}{\partial \varphi}$ is absent in the expression  \eqref{defrange2} used in \cite{Baume:2016psm,Valenzuela:2016yny,Blumenhagen:2017cxt}. 
For eq.~\eqref{defrange2} to be a good approximation to the allowed field range, we need to ensure 
\be
G_{\alpha \beta}\left(\varphi\right)\frac{\partial\overline{\phi}^{\alpha}}{\partial \varphi}\frac{\partial\overline{\phi}^{\beta}}{\partial \varphi} \leq \varepsilon G_{\varphi \varphi} \left(\varphi\right)\, ,
\label{validity}
\ee
where we can think of $\varepsilon$ as a measure of the accuracy of the approximation or, in other words, a measure of how well do the fields that minimize the potential 
solve the eoms \footnote{ Note that $\varepsilon$ will be vanishing only in the case of consistent decoupling between sectors.}. If this condition cannot be satisfied along the whole trajectory, there must exist a critical value, $\varphi_{\text{app}}$, for the displaced field beyond which the approximation fails. We define $\varphi_{\text{app}}$ as the solution to the constraint \eqref{validity} and, as we will illustrate, it will depend on the relative mass hierarchy. For $\hat{\varphi} > \hat{\varphi}_{\text{app}}$ 
significant corrections to the field range are expected and 
a multi-field analysis is necessary to capture the dynamics. 
For $\hat{\varphi} < \hat{\varphi}_{\text{app}}$,  the single-field approximation is valid.
%
%
The approximation for the field range \eqref{defrange2}
is, then, valid within a ball in the moduli space  centered at the minimum of the potential 
with
radius $\hat{\varphi}_{\text{app}}$. 
The dynamical field range 
is thus
valid  within this approximation up to $\hat{\varphi}_{c} = \text{min}\left\lbrace\hat{\varphi}_{\text{app}},\hat{\varphi}_{\text{back}}\right\rbrace$.
We will check that in the absence of mass hierarchies,
$\hat{\varphi}_{\text{app}} < \hat{\varphi}_{\text{back}}$.  The real cutoff of the effective theory should then be obtained by solving the full multi-field dynamics and, thus, the field range obtained in \cite{Baume:2016psm,Valenzuela:2016yny,Blumenhagen:2017cxt} will differ from the dynamical field range. 
On the contrary, the 
allowed field range is given by $\hat{\varphi}_{\text{back}}$ 
only if a sufficiently large mass hierarchy is assured. In this regime, corrections to the action
generically flatten the potential \cite{Dong:2010in}. Then, one may quantify the backtreaction by perturbing the heavy fields around the minimum of the potential. This procedure has been systematically analyzed for $\mathcal{N}=1$ supergravity \cite{Buchmuller:2015oma} (see \cite{Landete:2016cix} for a shortcut).

\subsection{Is the trajectory a geodesic?}

 We are now in a position to address whether the trajectory which defines the dynamical field range is a geodesic. Recent studies of swampland distances, see, e.g., \cite{Grimm:2018ohb,Blumenhagen:2018nts}, relied on geodesic trajectories in field space for massless axions. We will critically assess under what conditions this assumption holds for massive axions.
In multi-field setups, it is convenient to define a unit vector tangent to the trajectory which for the 
present case is:
%
\be
T^{a} := \frac{ \dot{\phi}_{0}^{a}}{\dot{\phi}_{0}} \rightarrow \dot{\phi}_{0}T^{\varphi} = \dot{\varphi}\, , \, \dot{\phi}_{0}T^{\alpha} = \frac{\partial \overline{\phi}^{\alpha}}{\partial \varphi} \dot{\varphi}\, .
\label{Ta}
\ee
%
 As in 
 \cite{GrootNibbelink:2001qt}, one may define an orthonormal basis in field space by taking covariant derivatives of the tangent vector, $D_{t}T$. For $\dot{\phi}_{0}\neq 0$, the trajectory obtained by solving the eoms \eqref{eoms} will be a geodesic in field space if $D_{t}T^a= 0$. Then, a sufficient condition for the trajectory considered to be a geodesic is 
\be
\left(\frac{\partial \overline{\phi}^{\alpha}}{\partial \varphi}\right) \approx 0 \rightarrow \left|D_{t}T\right| \approx 0\, .
\label{geodesic}
\ee
%
Indeed, one may approximate the trajectory by a geodesic  when eq.~\eqref{validity} is satisfied. 
This 
is a reasonable approximation if there are large mass hierarchies, and in 
the
absence of non-trivialities coming from the geometry of the field space (see, e.g., \cite{Achucarro:2010jv} for the latter point). Note that the trajectories studied in \cite{Baume:2016psm,Valenzuela:2016yny,Blumenhagen:2017cxt} are indeed geodesics since \eqref{geodesic} is assumed.

 


\section{ An Illustrative example}
\label{sec:toymodel}

We now try to quantify the statements made in the previous section by analyzing a single-field inflationary model based on D7-branes in Type IIB flux compactification of a toroidal orientifold. We refer the reader to the appendices for a similar discussion for closed-string models. For ease of comparison, we will focus on an example used in \cite{Blumenhagen:2017cxt} which involves non-geometric fluxes, we refer the reader to that reference for more details. The inflaton candidate, $\varphi$, is the axionic component of the position modulus of a D7-brane. The $\mathcal{N}=1$ supergravity Lagrangian describing this model is given by the F-term potential where:
\bea
K & = & -\log\left[\left(S+\bar{S}\right)\left(U+\bar{U}\right) - \frac{1}{2}\left(\Phi + \bar{\Phi}\right)^2\right] \nonumber\\ 
& - & 2\log\left(U+\bar{U}\right) - 3\log\left(T+\bar{T}\right) \, , \\
\label{kop}
W & = & \mathfrak{f}_{0} + 3 \mathfrak{f}_{2}U^{2} - hSU -qTU - \mu \Phi^2 \, .
\label{supo}
\eea
One may find a tachyon-free non-supersymmetric AdS vacuum with the following mass hierarchies \footnote{As a disclaimer, for the sake of simplicity we will consider 
a simplified (though less realistic) constant uplift term. However, we have checked that the final results do not differ quantitatively if one considers, for instance, an F-term uplift coming from a nilpotent goldstino \cite{Ferrara:2014kva}.} 
\be
M^{2}_{\text{closed}, \chi} \sim \alpha_{i}\frac{hq^{3}}{\mathfrak{f}_{0}^{\frac{3}{2}}\mathfrak{f}_{2}^{\frac{1}{2}}}, \, \, M^{2}_{\varphi} \sim \mu \frac{q^{3}}{\mathfrak{f}_{0}^{\frac{3}{2}}\mathfrak{f}_{2}^{\frac{1}{2}}} \, ,
\ee
where $\chi$ denotes the saxionic partner of $\varphi$, closed refers to the mass scale of all the closed string sector moduli and $\alpha_{i}$ are numerical factors. In this example, as we vary $\varphi$, the vevs for the rest of the moduli which minimize the scalar potential are
\begin{small}
	\bea
	\overline{s} &\sim& \frac{2^{7/4}3^{1/2}}{5^{1/4}}\frac{\left(\mathfrak{f}_{0} + \mu \varphi^2\right)^{1/2}\mathfrak{f}_{2}^{1/2}}{h} \, ,  \overline{u} \sim \frac{1}{10^{1/4}3^{1/2}}\left(\frac{\mathfrak{f}_0+\mu\varphi^2}{\mathfrak{f}_{2}}\right)^{1/2}\, , \nonumber \\
	\overline{t} &\sim& \frac{5^{3/4}3^{1/2}}{2^{1/4}}\frac{\left(\mathfrak{f}_{0} + \mu \varphi^2\right)^{1/2}\mathfrak{f}_{2}^{1/2}}{q} \, ,
	\label{sols}
	\eea
\end{small}
while the moduli not included in the above equation are stabilized at zero vev for all values of $\varphi$. Finally, we note that the mass hierarchy between the various sectors is given by
\be
\frac{M_{\varphi} }{M_{\text{closed}, \chi}} \sim \sqrt{\frac{\mu}{h}} \, .
\label{hierarchy}
\ee
%
Thus, in this example, the only way to generate a hierarchy is to tune the ratio \eqref{hierarchy}. 
It has been shown in 
 \cite{Bielleman:2016olv,Valenzuela:2016yny,Landete:2017amp} 
 that
 tuning $\mu \ll 1$ 
 can be done while satisfying the experimental constraints on the inflationary parameters given by the recent PLANCK data
 \cite{Ade:2015lrj}. 
 One should, however, be cautious about this tuning.\footnote{As argued in \cite{Blumenhagen:2017cxt}, this setup admits a four-fold description in the context of F-theory, and hence
 $\mu$ should be quantized since it arises from a $G_{4}$-flux. Thus, the mass ratio \eqref{hierarchy} cannot be tuned arbitrarily small. We would like to mention a possibility to overcome the  aforementioned problems. One may obtain a small mass ratio by placing the D7-brane in a warped region. This approach was used in \cite{Escobar:2015fda} for type IIA compactifications with D6-branes. In this case the mass ratio will be $\frac{M_{\varphi}}{M_{\text{closed}}}\sim\sqrt{\frac{\mu}{Zh}}$, where $Z$ is the warping felt by the D7-brane. By tuning $Z$, one may obtain a sufficiently small mass ratio while $\mu$ and $h$ are quantized.}
As our discussions below depend only on the existence of a mass hierarchy and not on the details of how it is achieved, we will sidestep these challenges and focus on exploring the consequences.
Nonetheless, we should stress that generating mass hierarchies in stabilized flux compactifications remains one of the major challenges in string theory (see \cite{Blumenhagen:2014nba} for a discussion along these lines), and may well provide a key ingredient in constructing controlled models with super-Planckian field ranges. In the following, we will check the validity of the approach taken by the RSDC in this concrete case.
\subsection{Validity of the approach}

To consistently ignore the kinetic terms of the heavy moduli,  we should check if eq. \eqref{validity} is satisfied. For a toroidal compactification this is translated into
\be
\frac{1}{G_{\varphi \varphi}\left(\varphi\right)} \sum_{\alpha}\beta_{\alpha}\left(\frac{\partial}{\partial \varphi} \log \left(\overline{\phi}^{\alpha}\right)\right)^2 \leq \varepsilon\, .
\label{toroidalvalidity}
\ee
%
Then, substituting the minimization condition \eqref{sols} into the former expression, one arrives at
\be
7\frac{\hat{\varphi}^2}{\hat{\varphi}^2 +\left(\frac{M_{\text{closed}}}{M_{\varphi}}\right)^2} \lesssim  \varepsilon \left(\frac{M_{\text{closed}}}{M_{\varphi}}\right)^2\, .
\label{torviab}
\ee
The critical value where the approximation breaks down, $\hat{\varphi}_{\text{app}}$, may be obtained by solving the above expression. We observe that \eqref{torviab} implies $\left(\frac{\partial \overline{\phi}^{\alpha}}{\partial \varphi}\right) \lesssim \sqrt{\varepsilon} \left(\frac{M_{\text{closed}}}{M_{\varphi}}\right)$. From \eqref{torviab} we see explicitly, as we previously anticipated, that there are two limiting regimes: 
\begin{enumerate}
	\item[1.] Absence of a mass hierarchy
\be
\frac{M_{\text{closed}}}{M_{\varphi}}\sim 1 \rightarrow \hat{\varphi}_{c}=\hat{\varphi}_{\text{app}}\sim \sqrt{\frac{\varepsilon}{7}}\,M_{\text{Pl}} < 1 M_{\text{Pl}}\, ,
\label{nohie}
\ee
the critical field range is always subplanckian and signals the breakdown of the approximation  eq.~\eqref{toroidalvalidity} rather than the breakdown of the EFT.
\item[2.] Presence of a large mass hierarchy
\be
\left(\frac{M_{\text{closed}}}{M_{\varphi}}\right)^2 > \frac{7}{\varepsilon} \rightarrow \hat{\varphi}_{c} = \hat{\varphi}_{\text{back}} \sim \frac{M_{\text{closed}}}{M_{\varphi}} M_{\text{Pl}}\, ,
\label{largehie}
\ee
eq. \eqref{validity} is satisfied for all $\varphi$ and thus the approach used in the RSDC is valid. The critical field range is then super-Planckian.
\end{enumerate}
Here, we point out that  in \eqref{toroidalvalidity} the backreaction effect of each modulus is weighted by its mass compared to that of the axionic field. In this example, all the stabilized moduli have roughly the same mass and, thus, their effects contribute on equal footing. 
In a KKLT-like \cite{Kachru:2003aw} or LVS \cite{Balasubramanian:2005zx} compactification, 
however, there is a hierarchy of scales 
among the closed-string sectors (see \cite{Hebecker:2014eua,Ibanez:2014swa} for examples). 
In this case, treating all variables dynamically, the RSDC constraint \eqref{defrange2} is weaker compared to the constraint imposed by destabilization of the K\"ahler modulus when a large mass hierarchy between the axion and the complex structure sector is assured. Nonetheless, only by using \eqref{validity} one may find the needed mass hierarchy between the axion and the K\"ahler sector to trust the single-field approximation.


\subsection{The RSDC in different regimes}

\begin{figure}[h]
	\includegraphics[width=7.8cm]{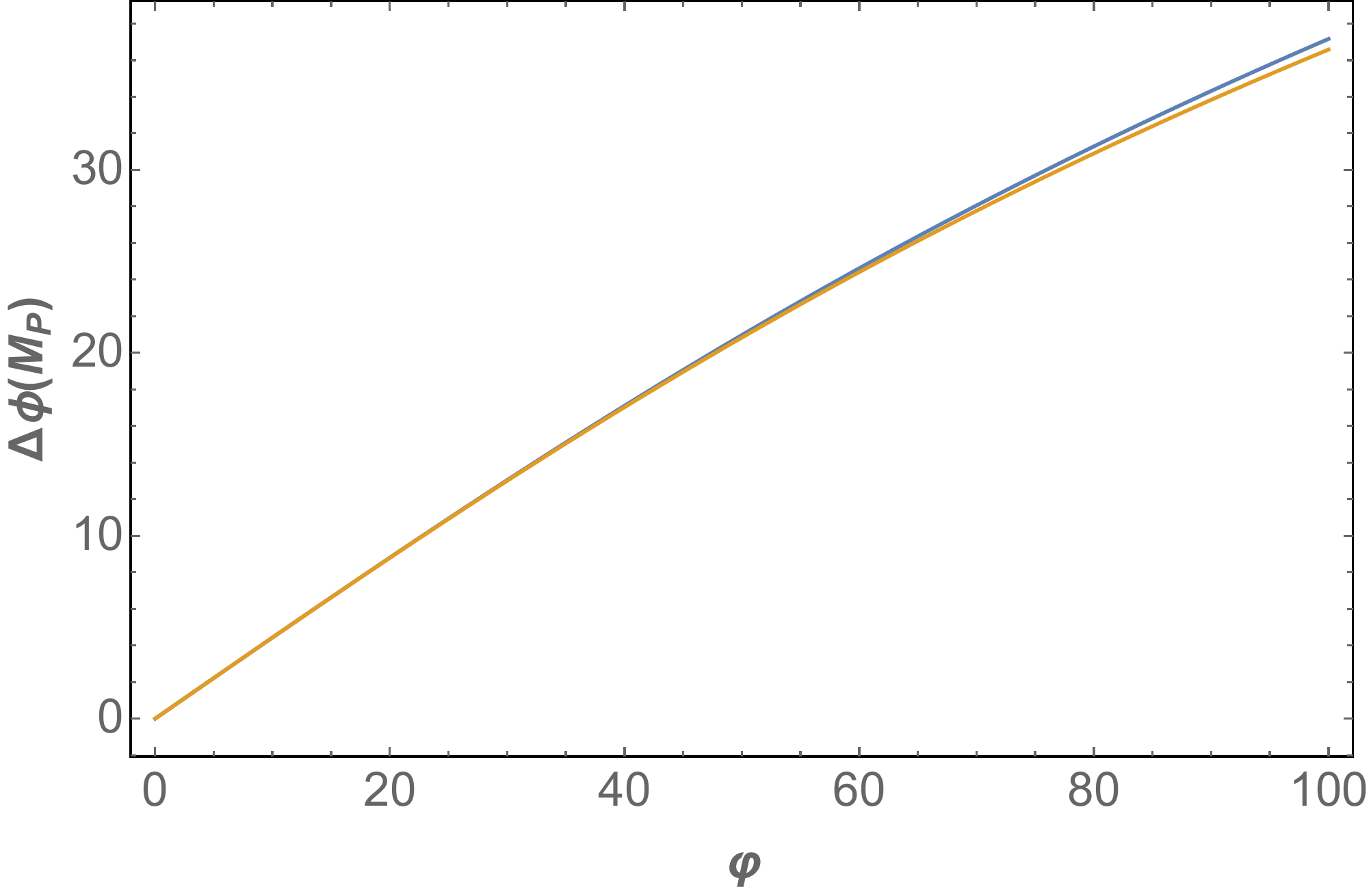}
	\centering
	\caption{Field ranges obtained by \eqref{rangebetter} (blue) and \eqref{defrange2} (orange) for $\frac{\mathfrak{f}_{0}}{\mu} \sim 10^{4}$ and $G_{\varphi \bar{\varphi}}|_{0} \sim 2\cdot10^{-1}$.}
	\label{fig:comp2}
\end{figure}

\paragraph{Large mass hierarchies.}
We have shown in \eqref{largehie} that when $M_{\varphi} \ll M_{\text{closed}}$ the dynamical field range is approximately measured along a geodesic trajectory and thus
\be
\Delta\phi  =  \int \sqrt{G_{ab}\dot{\phi}^{a}\dot{\phi}^{b} }dt \approx \int \sqrt{G_{\varphi \varphi}\left(\varphi\right)}d\varphi \sim \frac{M_{\text{closed}}}{M_{\varphi}}\log\left(\varphi\right)\, .
\label{fieldrange2}
\ee
The critical field range is then $\hat{\varphi}_{c} = \hat{\varphi}_{\text{back}} \sim \frac{M_{\text{closed}}}{M_{\varphi}} M_{\text{Pl}}> 1 M_{\text{Pl}}$, in agreement with \cite{Blumenhagen:2017cxt}. We see from Figure \ref{fig:comp2} that the dynamical field range indeed agrees with the one measured along the geodesic trajectory.
\paragraph{No mass hierarchies.}

\begin{figure}[h]
	\includegraphics[width=7.8cm]{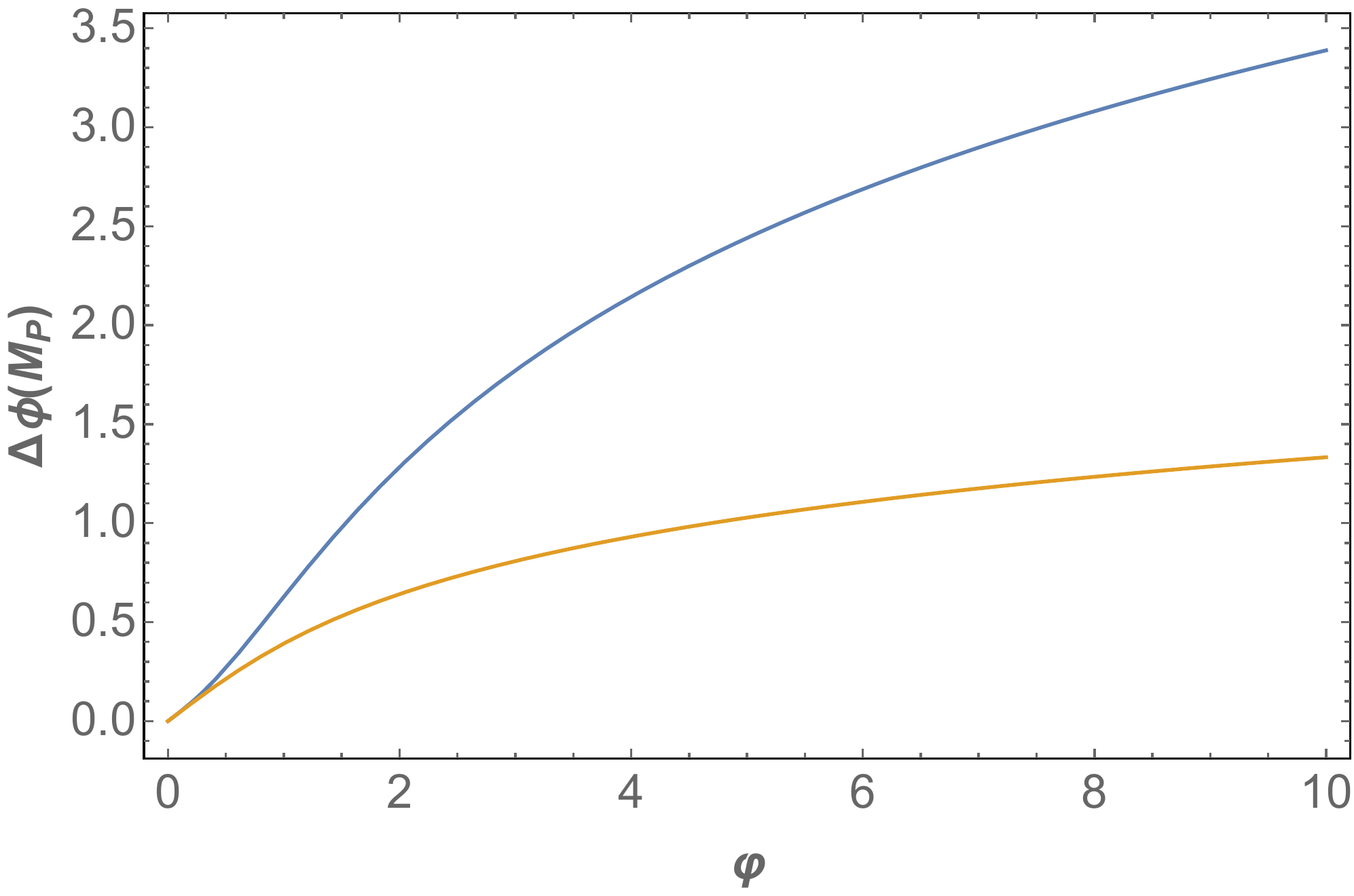}
	\centering
	\caption{Field range obtained by \eqref{rangebetter} (blue) and \eqref{defrange2} (orange) for $\frac{\mathfrak{f}_{0}}{\mu} \sim 1$ and $G_{\varphi \bar{\varphi}}|_{0} \sim 2\cdot10^{-1}$. In this case $\hat{\varphi}_{\text{app}} \sim 0.1 M_{\text{Pl}}$ for $\varepsilon = 0.1$}
	\label{fig:comp}
\end{figure}

We now compare the RSDC estimate of the field range using eq.~\eqref{defrange2} with the measure eq.~\eqref{rangebetter} in cases where there are no mass hierarchies outside the validity regime we have established in \eqref{validity}. 
We have shown in \eqref{nohie} that this approximation will be valid until $\hat{\varphi}_{\text{app}}$. We show in Figure \ref{fig:comp} how the two field ranges deviate from each other beyond this point. Then, in this regime one should solve numerically the full multi-field dynamics. 
We have also computed numerically the field range with the actual solution to the equations of motion and found that the behavior depicted in Figure \ref{fig:comp} 
does not change qualitatively.
We observe that, asymptotically, for large $\varphi$ displacements, the dynamical field range grows at best logarithmically, in agreement with the SDC. Note that the kinetic terms of the minimized fields are sizable, precisely, when there is no mass hierarchy, i.e. , $\mu \sim h$ and, as anticipated, our computation shows that the larger the backreaction effects the larger the deviation between the naive approximate \eqref{fieldrange2} and the dynamical field range.
%

\section{Discussion}

In this letter we have analyzed carefully the viability of the constraints imposed by the RSDC on transplanckian field displacements of massive axions.
We illustrated our general results with axions in type II compactifications.

One of our findings is that minimizing the scalar potential without taking into account peculiarities of the field space metric and the mass hierarchies involved may lead to inconsistent results. 
In the examples analyzed in the RSDC literature, the available field range is measured along a geodesic trajectory while neglecting the kinetic terms of the other fields. This is, indeed, a reasonable approximation if there is a large mass hierarchy and when the backreaction effects are mild. Precisely, in this approximation, the critical field range is extended to a super-Planckian value.
 In  the absence of a mass hierarchy, the approximation breaks down before the onset of the logarithmic behavior that defines the RSDC field range and, then, the sub-Planckian field restriction imposed by the RSDC cannot be trusted. This fact points to one of our premises, which is the importance of analyzing the dynamical field range. In order to constrain field displacements by means of the SDC, it should be applied to physical trajectories. Our findings support this conjecture since the logarithmic behavior of the field range holds asymptotically in all the cases studied, but the concrete point where the EFT suffers a breakdown can only be obtained explicitly by solving the eoms. Indeed, we have proved that in the regime where minimizing the scalar potential is a good approximation to solving the eoms,
  the dynamical field range is well approximated by the geodesic distance. Beyond that regime, the prescription given by the RSDC loses its physical validity and a multi-field analysis is needed. 

The above subtleties with multi-field dynamics have appeared in the study of inflation.
For example, it is well-known that the kinetic terms of the heavy fields can lead to turns in field space. As a result, minimizing the scalar potential for the heavy fields as we vary the light field (the inflaton in this case) does not in general solve
the eoms. The deviation of the true trajectory from a geodesic can be parametrized 
by $\eta_{\perp}$ \cite{Shiu:2011qw,Achucarro:2012yr,Tolley:2009fg,Rubin:2001in}. 

Given the role of mass hierarchies in defining the dynamical field range, it is important to explore new ways of generating such hierarchies in string compactifications. 
While ideas such as employing large ratios of flux quanta have been put forth, they seem to be challenging 
to implement in practice, as can be seen in Type IIB settings with non-geometric fluxes \cite{Blumenhagen:2014nba}.
 It would be interesting to explore other proposals such as the use of warping for the open-string sector \cite{Escobar:2015fda} 
 and the use of $g_{s}$ loops for K\"ahler moduli axions \cite{Cicoli:2014sva}. Finding new ways to generate mass hierarchies in string constructions may hold the key to extending the dynamical field range in quantum gravity. We leave for future work the extension of these findings to generic Calabi-Yau manifolds.
\vspace{1.0cm}

{\bf Acknowledgements. }We thank Fernando Marchesano and Wieland Staessens for discussions. This work is supported in part by the DOE grant DE-SC0017647 and the Kellett Award of the University of Wisconsin.

\bibliography{finalRSC}

\appendix

\section{Appendix: Refined Swampland Conjecture for Closed String Models}

For completeness we
consider models involving only closed-string moduli.
We will show that in absence of a mechanism to create a hierarchy between an axionic direction and the rest of the closed string sector, we obtain similar results 
as in the $\mu \sim \h$ case
analyzed in the main text. 
Along the same lines, we will see that the available field range 
estimated by the RSDC is not valid with large displacements of the axionic field. 
The approximation breaks down before the onset of the logarithmic behavior of the field range, i.e., $\hat{\varphi}_{\text{app}} < \hat{\varphi}_{\text{back}}$.

For example, in \cite{Baume:2016psm} 
the displaced axionic field considered is a linear combination of complex structure axions in 
Type IIA compactifications.
In the examples analyzed, there is no way to create a mass hierarchy between the closed string sector and the displaced field, then,
$\hat{\varphi}_{c}$ was found to be ``flux-independent" ( similar examples were analyzed in
\cite{Valenzuela:2016yny,Blumenhagen:2017cxt}).

As an illustrative example one may consider the following toroidal compactification exemplified in \cite{Blumenhagen:2015kja}
\begin{small}
\bea
K & = &  -\log\left[\left(S+\bar{S}\right)\right] -3\log\left(U+\bar{U}\right) - 3\log\left(T+\bar{T}\right)\, , \\
W & = & -\mathfrak{f}_{0} - 3 \mathfrak{f}_{2}U^{2} - hSU -qTU \, ,
\label{kop2}
\eea
\end{small}
where we consider the following massive axion
\be
\varphi = hs_{2} + q t_{2} \, ,
\ee
Here, the subscript 2 denotes the axionic component of the corresponding superfield and the subscript 1 denotes the saxionic component.

In this example, 
all modulli are stabilized in a tachyon-free AdS vacuum with the moduli masses given by
\be
M_{\text{mod},i} = \alpha_{i}\frac{hq^{3}}{\mathfrak{f}_{0}^{\frac{3}{2}}\mathfrak{f}_{2}^{\frac{1}{2}}} \, ,
\ee
where $\alpha_{i}$ are order one numerical factors. 
The available field range
given by \eqref{defrange2} 
is
\bea
\Delta \phi & = &  \int \left(h^{2}K^{S\bar{S}}\left(\varphi\right) + q^{2}K^{T \bar{T}}\left(\varphi\right)\right)^{-1/2}d\varphi  \\
& = & \int \sqrt{\frac{1}{\frac{4}{3}q^{2}t_{1}^{2}\left(\varphi\right) + 4h^{2}s_{1}^2\left(\varphi\right)}}d\varphi := \int \sqrt{G_{\varphi \varphi}\left(\varphi\right)}d\varphi \, . \nonumber
\label{naiveclosed1}
\eea
%
This
expression is valid as long as backreaction effects are not sizable.
%
Including the kinetic terms of moduli whose stabilized values depend on $\varphi$, $\Delta \phi$ is modified to
\begin{small}
	\be
	\int \sqrt{G_{\varphi \varphi}+\left(\frac{\partial u_{1}}{\partial \varphi}\right)^2K_{U \bar{U}} + \left(\frac{\partial s_{1}}{\partial \varphi}\right)^2K_{S \bar{S}} + \left(\frac{\partial t_{1}}{\partial \varphi}\right)^2K_{T \bar{T}}} d\varphi 
	\label{naive2}
	\ee
\end{small}
%
The approximation 
of
neglecting the kinetic terms of the other fields
sets the range of validity
$\varphi_{\text{app}}$ which depends on the mass hierarchy between the axionic displaced field, $\varphi$, and the different moduli which are backreacting. 
While we do not have an analytic expression for the fields minimizing the scalar potential, we can evaluate the above expression for $\Delta \phi$ numerically.
In this 
example, 
there is no mass hierarchy and, as a result, the maximum displacement, within this approximation, for the canonically normalized field is flux-independent and is 
equal
to
$\hat{\varphi}_{app} = 0.056 M_{\text{Pl}}$ for $\varepsilon = 0.1$
This critical field range may be delayed for the non-canonically normalized field by making $\mathfrak{f}_{0}$ large, while it is constant for the canonically normalized value as observed in \cite{Baume:2016psm}.

\begin{figure}[h]
	\includegraphics[width=7.8cm]{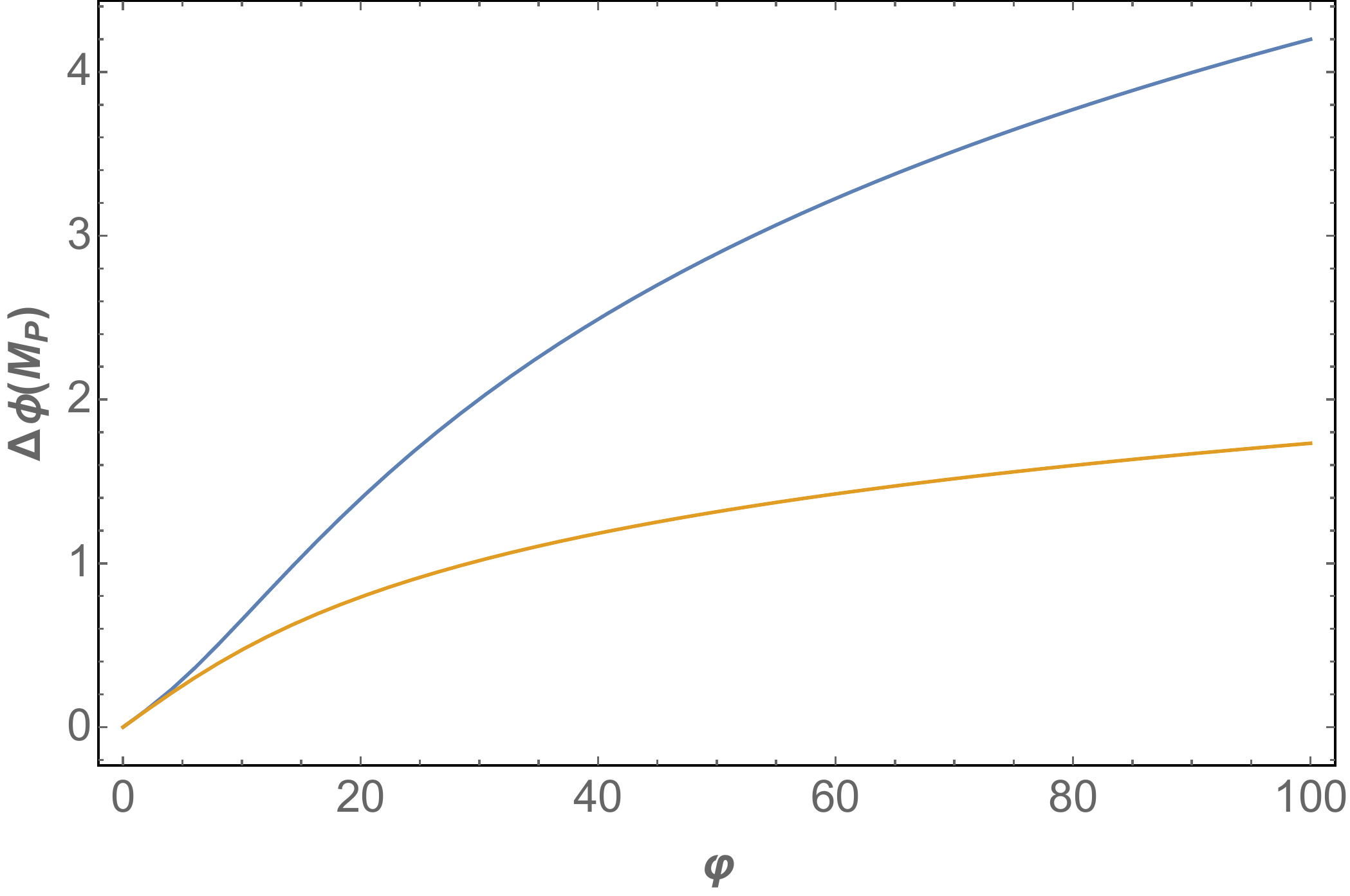}
	\centering
	\caption{Field ranges obtained by \eqref{naive2} (blue) and \eqref{naiveclosed1} (orange) for a stabilization with $h=1, \, q=1, \, \mathfrak{f}_{0} = -2, \mathfrak{f}_{0} = -4, $.}
	\label{fig:comp3}
\end{figure}

Finally, we compare in Figure \ref{fig:comp3} the allowed field ranges outside the validity regime 
with and without including the kinetic terms of the rest of the fields.
%
%

In comparison, the allowed field range estimated by the RSDC is also "flux-independent" but is equal to $\hat{\varphi}_{\text{back}} \sim 0.7 M_{\text{Pl}}$ which is bigger than $\hat{\varphi}_{app}$. Thus, the
RSDC  neglects important corrections to the field range as we can see explicitly in Figure \ref{fig:comp3}.

\end{document}